\documentstyle[prb,aps,twocolumn]{revtex}
\begin{document}
\draft

\title{Pinned states in Josephson arrays: A general stability theorem}
\author{Mauricio Barahona$^{\ast}$ and Steven H. Strogatz$^{\dagger}$}
\address{$^{\ast}$Ginzton Laboratory, Stanford University, Stanford, CA 94305}
\address{$^{\dagger}$Department of Theoretical and Applied Mechanics and
Center for Applied Mathematics, Kimball Hall, Cornell University, Ithaca,
NY 14853}
\date{\today}
\maketitle

\begin{abstract}
Using the lumped circuit equations, we derive a stability criterion for
superconducting pinned states in two-dimensional arrays of Josephson
junctions.  The analysis neglects quantum, thermal, and inductive effects,
but allows disordered junctions, arbitrary network connectivity, and
arbitrary spatial patterns of applied magnetic flux and DC current
injection.  We prove that a pinned state is linearly stable if and only if its
corresponding stiffness matrix is positive definite.  This algebraic
condition can be used to predict the critical current and
frustration at which depinning occurs.
\end{abstract}


Collective pinning occurs in a wide variety of coupled physical
systems. Examples include vortices in Type-II
superconductors, cracks and dislocations in solids, and charge-density
waves in quasi-one-dimensional metals.~\cite{physdep}
In each case, when the system is subjected to an external
constant drive, it remains motionless until the drive
exceeds a critical value (the depinning threshold) after which the system
begins to move.  The pinning is collective in the sense that it involves
interactions among many coupled subsystems, typically in the presence of
disorder.  Hence it is often difficult to predict the depinning threshold
theoretically.

Here we study collective pinning for a relatively tractable class of model
systems: two-dimensional (2D) arrays of Josephson junctions.  Besides their
technological applications,~\cite{scbooks} Josephson arrays can be used to
explore fundamental questions in statistical mechanics (such as phase
transitions), and in nonlinear dynamics (such as synchronization and
spatiotemporal pattern formation).~\cite{generalphys} In addition, they
have been proposed as clean models for layered and granular high-T$_{\rm c}$
superconductors.~\cite{kleiner,2Dnum}
As such, their depinning could be relevant to the
understanding of the onset of resistance in the current-voltage
characteristics of high-T$_{\rm c}$ samples.~\cite{mpa}

Several advances have occurred recently in the
numerical~\cite{2Dnum,stroud,joel,dominguez} and
analytical~\cite{2Danalytical,RS,ladder} investigation of 2D Josephson
arrays, thanks in part to an influx of ideas from nonlinear dynamics.
In this paper, we analyze depinning in 2D arrays from this perspective.
Using a compact matrix notation, we show that the linear stability problem
for pinned states can be mapped onto the classical mechanical problem of
small oscillations in a network of coupled, damped linear oscillators.
The results apply to 2D arrays of any given topology.  There are also no
restrictions on the capacitances, resistances, or critical currents of the
junctions, nor on the spatial patterns of DC current injection and applied
magnetic flux.  Our main result is that a pinned state is stable if and
only if its corresponding stiffness matrix $K$ is positive definite.
This matrix $K$ changes with the pinned configuration and depends
on the connectivity and disorder of the array.  A corollary is that any
pinned state with all phases $|\phi_i| < \pi/2$ is guaranteed to be stable.
We also prove that depinning can never occur via a Hopf bifurcation; only
zero-eigenvalue bifurcations are possible.

Our analysis is based on several simplifying assumptions.  First, we
neglect thermal fluctuations; that is, we assume zero temperature. Second,
we assume that the superconducting islands in the array are
large enough that quantum (charging) effects are negligible. Thus,
the phase $\theta_i$ of the complex macroscopic wavefunction at each island
is a well-defined classical variable.  Third, we assume that the junctions
between islands are small enough that they can be approximated as lumped
elements. Therefore, the junction
between two islands $\ell$ and $m$ can be described by a point
gauge-invariant phase difference
\begin{equation}
\phi_i = \theta_{\ell} - \theta_m -
\frac{2 \pi}{\Phi_0} \int_{\ell}^{m} {\bf A} \cdot d{\bf l}
\label{eq:phasedif}
\end{equation}
where ${\bf A}$ is the total magnetic vector potential and
$\Phi_0=h/(2e)$ is the quantum of magnetic flux.
Fourth, we model each junction by the standard RCSJ equivalent
circuit~\cite{scbooks,steve} with superconducting, resistive, and
capacitive channels in parallel.
Then the junction dynamics obeys a damped driven pendulum equation
\begin{equation}
\mu_i \ddot{\phi_i} + \gamma_i \dot{\phi_i} +
\eta_i \sin \phi_i = i_i^b
\label{eq:RCSJ}
\end{equation}
with effective mass $\mu_i=\Phi_{0} C_i/(2 \pi I_{c0})$,
damping $\gamma_i=\Phi_{0}/(2 \pi R_i I_{c0})$,
and restoring strength  $\eta_i=I_{ci}/I_{c0}$.  The capacitance
$C_i$, resistance $R_i$, and critical current $I_{ci}$ are fabrication- and
material-dependent parameters that
characterize junction $i$.
The drive is given by the normalized current $i_i^b$, measured in units of
$I_{c0} = \langle I_{ci} \rangle$, the average critical current of the
junctions in the array.

In dealing with arrays, it is useful to introduce a
vector-matrix notation,~\cite{joel,dominguez,RS,strang}
where the variables are now vectors of three types:
node vectors of dimension $n$ (e.g.\ $\theta$);
edge vectors of dimension $e$ (e.g.\ $\phi$ and $i^b$); and
cell vectors of dimension $c$ defined at each plaquette.
(More precisely, $n$ is the number of independent nodes, after one
node is grounded and taken as reference.~\cite{strang})
The edge and node variables are related through an $e \times n$ edge-node
connectivity matrix $A$ that encodes the topology of the array, including
its boundary conditions such as the presence (or absence) of edges.
Similarly, an $e \times c$ edge-cell matrix $B$ transforms between edge and
cell variables, 
in what amounts to taking a discrete curl.

Within this framework, the nonlinear constitutive law~(\ref{eq:RCSJ}) can
be compactly written as
\begin{equation}
\mu \ddot{\phi} + \gamma \dot{\phi} + \eta \sin \phi = i^b
\label{eq:RCSJvec}
\end{equation}
where $\mu={\rm diag}(\mu_i)$, and $\gamma$ and $\eta$ are similarly
defined diagonal matrices.
Each junction is allowed to have a different capacitance, resistance, and
critical current, as recorded in the matrices $\mu$, $\gamma$, and $\eta$.

When junctions are interconnected to form a network, there exist
topological constraints which can be expressed in terms of the
connectivity matrices $A$ and $B$.
First, 
the currents
must satisfy Kirchhoff's current law~\cite{strang}
\begin{equation}
A^T i^b = i^{\rm ext},
\label{eq:KCL}
\end{equation}
where the vector $i^{\rm ext}$ gives the balance of normalized current
at each node, and reflects the
particular scheme of current injection/extraction for each experimental
device. For instance, in the usual experimental setup, where a  uniform DC
current $I_{\rm dc}$ is injected (extracted) at the bottom (top) nodes, all
the components of
$i^{\rm ext}$ will be zero except those at the bottom (top) boundary,
which will be equal to $I_{\rm dc}$ ($- I_{\rm dc}$).
Our analysis, however, is valid for an arbitrary
injection scheme, as long as the bias currents are time-independent.

The second topological constraint is the flux quantization in each
cell of the array. We assume the
simplest case where all self-fields due to inductance effects are
neglected.  Then the flux
quantization is given by
\begin{equation}
B^T \phi + 2 \pi F = 2 \pi \zeta \equiv 0,
\label{eq:FQ}
\end{equation}
where $\zeta$ is a cell vector of integers (topological vorticities)
that have no dynamical relevance, and can be redefined as zero with
no loss of generality.~\cite{RS}
The cell vector $F$ records
the external flux through each plaquette, measured in units of the flux
quantum.
In experiments, the external magnetic field is often spatially uniform
across the array. Then $F$ is a constant vector
with value $f=\Phi_{\rm ext}/\Phi_{0}$.
Our analysis holds more generally for any time-independent spatial pattern
of applied flux.

For the no-inductance case assumed here, the
transformation~(\ref{eq:phasedif})
between junction and
island phases is given in vector form by
\begin{equation}
\phi= A \theta - \varphi
\label{eq:phitheta}
\end{equation}
where $\varphi$ is a time-independent edge flux vector,
fixed by our choice of gauge but subject to
$B^T \varphi = 2 \pi F$,
which follows directly from~(\ref{eq:FQ}) noting that $B^T A \equiv 0$,
from the definition of the topological matrices.~\cite{strang}
From~(\ref{eq:RCSJvec}),~(\ref{eq:KCL})  and~(\ref{eq:phitheta}), we obtain
the governing vector equation of the system:
\begin{equation}
A^T \mu A \; \ddot{\theta} + A^T \gamma A \;\dot{\theta} +
A^T \eta \; \sin(A \theta - \varphi) = i^{\rm ext}.
\label{eq:dynamvec}
\end{equation}

From now on, we focus on the pinned states of the array.  These correspond
to static configurations $\theta^{\ast}$ of~(\ref{eq:dynamvec}),
given implicitly by
\begin{equation}
A^T \eta \, \sin ( A \theta^{\ast} - \varphi) = i^{\rm ext}.
\label{eq:fixedpoint}
\end{equation}
Typically this nonlinear algebraic system~(\ref{eq:fixedpoint}) has
multiple solutions.  Each solution depends parametrically on the external
current vector $i^{\rm ext}$ and the applied flux vector $F$. (In the usual
experimental setup, these are determined by the scalars $I_{\rm dc}$ and
$f$, respectively.)
As $i^{\rm ext}$ or $F$ are varied,
the linear stability of a given static
configuration $\theta ^{\ast}$ can change.
This signals the transition
to another state of the system.  If the new state is still pinned,
the transition corresponds to a static rearrangement of phases and currents;
on the other hand, if the new state is
time-dependent, it corresponds to depinning and the onset of resistance.
(Because our analysis is local, it cannot distinguish between these two
types of transitions.)

To study the stability of the pinned states, let
$\theta = \theta^{\ast} + \alpha$
where $\alpha$ is a small perturbation.  Linearizing~(\ref{eq:dynamvec})
about $\theta^{\ast}$ yields
\begin{equation}
M \ddot{\alpha} + G \dot{\alpha} + K \alpha = 0,
\label{eq:linearsyst}
\end{equation}
where
\begin{equation}
M = A^T \mu A, \quad G = A^T \gamma A, \quad K = A^T \eta C^{\ast} A
\end{equation}
are the mass, damping, and stiffness matrices, respectively, and
\begin{equation}
C^{\ast} = {\rm diag} ( \cos \phi_i^\ast)
\label{eq:definitionC}
\end{equation}
is a diagonal matrix of the cosines of the phases of the
given static configuration.
Both $M$ and $G$ are symmetric, positive
definite matrices, since $A$ is a topology matrix and $\mu$ and $\gamma$
are diagonal matrices with positive masses and damping coefficients on the
diagonal.~\cite{strang}  However, $K$ is not necessarily positive
definite since the cosines on the diagonal of $C^{\ast}$ are not necessarily
positive.
We stress that the stiffness matrix $K$ is different
for each pinned state,
and it also changes parametrically with the externally tunable parameters.

Equation~(\ref{eq:linearsyst}) is familiar from the classical mechanical
problem of small oscillations in a network of coupled, damped harmonic
oscillators.~\cite{goldstein}  But the present stability problem is not as
trivial as it might seem.  Ordinarily one assumes that $K$ is positive
definite, but that need not be true here.  Also, recall that when damping
is present, normal modes cannot be used to decouple the system; in
mathematical terms, one cannot simultaneously diagonalize the three
symmetric matrices $M$, $G$, and $K$.  Therefore we
analyze~(\ref{eq:linearsyst}) from first principles.

A given pinned state is linearly stable if and only if the perturbation
$\alpha(t)$ decays to zero for all initial conditions.  Equivalently, all
the eigenvalues of~(\ref{eq:linearsyst}) must have strictly negative real
parts.  The characteristic equation
\begin{equation}
{\rm det} (\lambda ^2 M + \lambda G + K) = 0
\label{eq:eigenvalue2}
\end{equation}
cannot be solved explicitly, but one can still extract useful information
about the eigenvalues, as follows.  Suppose that~(\ref{eq:eigenvalue2})
holds for some $\lambda$.  Then there exists a (possibly complex)
eigenvector $x \ne 0$ such that
$\lambda ^2 M x + \lambda G x + K x = 0$.
Multiplying on the left by the complex conjugate transpose $x^{\dagger}$
yields
\begin{equation}
\lambda ^2 m + \lambda g + k = 0,
\label{eq:eigenvalue3}
\end{equation}
where $m= x^{\dagger} M x$, $g= x^{\dagger} G x$,
and $k= x^{\dagger} K x$
are scalars that depend on $x$.
Thus,
\begin{equation}
\lambda = \frac{ -g \pm \sqrt{g^2 -4 k m}}{2m}.
\label{eq:eigenvalue4}
\end{equation}
The key point is that $m>0$ and $g>0$ {\it for all} $x$, since  $M$ and $G$
are real and symmetric (hence Hermitian) positive definite matrices. On the
other hand, $K$ is not necessarily positive definite, so $k$ can have
either sign.
If $k > 0$, there are two subcases:
if $g^2-4 k m < 0$, the eigenvalues are complex conjugates with ${\rm
Re}(\lambda)  = -g/(2m) < 0$; otherwise the eigenvalues are both real and
negative.  In either case, the eigenvalues for $k>0$ lie in the left half
plane and therefore correspond to stable modes.
On the other hand, if $k < 0$, then
$\lambda_- < 0$, $\lambda_+ > 0$  so the $\lambda_+$ mode is unstable.
Finally, if $k = 0$, then
$\lambda_- < 0$, $\lambda_+ = 0$, and the $\lambda_+$ mode is neutral.

An important qualitative conclusion from these formulas is that any
eigenvalue of~(\ref{eq:eigenvalue2}) must be either pure real, or complex
with strictly negative real part.  In particular, pure imaginary
eigenvalues are forbidden.  An immediate consequence is that pinned states
can never undergo Hopf bifurcations; depinning can occur only through
zero-eigenvalue bifurcations~\cite{steve} such as saddle-node,
transcritical, and pitchfork bifurcations.

We now prove the main result: a pinned state is linearly stable if and only
if $K$ is positive definite.  To prove the ``if'' direction,
suppose that
$K$ is positive definite.  Then  $k>0$ for {\it all} eigenvectors $x$.
 From Eq.~(\ref{eq:eigenvalue4}) above, ${\rm Re}(\lambda)  < 0$ for all
$\lambda$ and, hence, the pinned state is linearly stable.

To prove the ``only if'' direction, it is equivalent to prove its
contrapositive, i.e., we assume that $K$ is not positive definite
and show that the pinned state is not linearly stable.
There are two cases.  If ${\rm det}(K) = 0$, then $\lambda = 0$
is a solution
of~(\ref{eq:eigenvalue2}), by inspection.  But $\lambda = 0$ corresponds to
a neutral mode, not a decaying mode as required for linear stability.
Next suppose  ${\rm det}(K) \ne 0$.
We outline a homotopy argument which proves
that~(\ref{eq:eigenvalue2}) has a root $\lambda > 0$.  The strategy is to
start with the undamped problem, where it is easy to show that
there is an unstable mode
if $K$ is not positive definite.
Then we continuously deform the undamped problem into
Eq.~(\ref{eq:eigenvalue2}), and show that the unstable eigenvalue {\it
remains} unstable throughout the deformation.  More precisely, consider the
one-parameter family of equations
\begin{equation}
{\rm det} (\lambda^2 M + p \lambda G + K) = 0
\label{eq:homotopy}
\end{equation}
where $0 \le p \le 1$ is a homotopy parameter.  At $p=0$,
Eq.~(\ref{eq:homotopy}) corresponds to an undamped system,
and normal modes
can be used to show explicitly that~(\ref{eq:homotopy}) has an
eigenvalue $\lambda(0) > 0$.  As $p$ varies continuously from 0 to 1, this
eigenvalue traces out a continuous curve $\lambda(p)$ in the complex plane.
The curve starts on the positive real axis since $\lambda(0) > 0$, and it
must stay there for all $p$ because any eigenvalue in the right half plane
must be pure real, as shown by~(\ref{eq:eigenvalue4}).  Moreover,
the curve cannot cross through the origin; from~(\ref{eq:homotopy}),
$\lambda(p) = 0$ for some $p$ would imply ${\rm det}(K) = 0$,
contrary to assumption.  Thus $\lambda(p) > 0$ for all $p$.  Setting
$p=1$ yields
the desired result that~(\ref{eq:eigenvalue2}) has a root $\lambda > 0$.

One consequence of this theorem is an implicit formula for the stability
threshold of a pinned state $\theta^{\ast}$.  As we vary the applied
current or magnetic field, $\theta^{\ast}$ and its associated matrix
$K$ will change.  The theorem implies that $\theta^{\ast}$ loses
stability precisely when
$K=  A^T \eta C^{\ast} A$ ceases to be positive definite.
This threshold is reached when the following
algebraic condition is satisfied for the first time:
\begin{equation}
{\rm det} (K) \equiv {\rm det} ( A^T \eta C^{\ast} A) = 0.
\label{eq:depinning}
\end{equation}
Hence the stability threshold for $\theta^{\ast}$
is determined {\it exclusively} by the array topology, by
the injection scheme and bias current (through $i^{\rm ext}$), by the
applied magnetic field $F$, and by the
disorder in the junctions' critical currents (via the matrix $\eta$). On
the other hand, it does not depend on the mass
(capacitance) and damping matrices $M$ and $G$. This means that overdamped
and underdamped systems have identical depinning thresholds.

Another corollary is that if
\begin{equation}
\cos \phi_i^{\ast} > 0, \;\; \forall i
\label{eq:lessthanpi/2}
\end{equation}
then that configuration is stable.
This follows from the fact that the diagonal matrix $\eta C^{\ast}$ of such a
configuration is positive definite; therefore $K$ is also positive
definite.~\cite{strang}
On the other hand, since $K$ can be positive definite even
if $C^{\ast}$ is not,
(\ref{eq:lessthanpi/2}) is only a {\it sufficient} (but not necessary)
condition for the stability of a pinned state.

The constraint~(\ref{eq:lessthanpi/2}) has a clear physical meaning for a
single, isolated Josephson junction.  Recall that as the bias current is
increased from zero, a single junction remains pinned until $ \phi  =
\pi/2$, at which point it depins to a running mode.~\cite{scbooks,steve}
Extrapolating naively from a single junction to an array, it is tempting to
conjecture that an array should depin when its ``most unstable" junction
first reaches $ \phi  =  \pi/2$.  Note, however, that this heuristic
depinning criterion is equivalent to ${\rm det} (C^{\ast})= 0$, rather
than the rigorous condition ${\rm det} (K) = 0$;  therefore, it is
not exact. Nevertheless, for the specific case of a ladder array with
square plaquettes and perpendicular current injection, we have shown
elsewhere~\cite{ladder} that it can provide a good approximation to the true
depinning threshold.

The algebraic condition~(\ref{eq:depinning}) can be used to ease the
numerical determination of the depinning threshold for 2D arrays.
For instance, the depinning current is usually obtained~\cite{stroud}
through dynamical simulations that resemble the actual experiment:
the current is ramped up adiabatically and the circuit differential equations
are numerically integrated until a running solution appears.
In contrast, we solve~(\ref{eq:fixedpoint}) and~(\ref{eq:depinning})
simultaneously to determine the critical
current and the bifurcating phase configuration as functions of all the
other parameters. This purely algebraic calculation
can be done by Newton's method or some other rootfinding scheme.
The results coincide with those found dynamically.~\cite{newdep}

Another theoretical approach to depinning uses thermodynamic
and quasistatic calculations of pinned states.~\cite{LAT,teitel,benedict}
One can show that the condition~(\ref{eq:depinning}) is
strictly equivalent to finding the point at which a given static
configuration ceases to be a minimum of the potential energy
\begin{equation}
V= - \theta^{\dagger} i^{\rm ext} - {\rm Tr} (\eta C).
\end{equation}
Thus a soft-mode condition~\cite{benedict}
rigorously predicts depinning, while the criterion based on
maximizing the quasistatic current induced by twisted boundary 
conditions~\cite{teitel} is only approximate.~\cite{stroud}
Note also that, although stable static configurations correspond to
local minima of $V$, we do not attempt here to obtain the {\it absolute}
minimum of the potential energy. This problem would require global
optimization methods, such as simulated annealing.

Our results open several promising lines of research.
First, our analytical framework facilitates exploration of
the effects of network connectivity on the depinning
of Josephson arrays. The implicit condition~(\ref{eq:depinning}) can
be turned into explicit, testable predictions of the
applied current and frustration at which depinning should occur.  It may be
possible to obtain analytical results for square and triangular
arrays of identical junctions, perhaps along the lines of recent work on
ladder arrays.~\cite{ladder}
Second, one should also try to take self-fields into account.  Preliminary
results suggest that the formulation given here can be generalized to
include inductance effects.~\cite{newdep}
Finally, it is important to study more quantitatively how disorder affects
the stability of pinned states, both as the inevitable result of
fabrication irregularities and as a design tool to manipulate the response
of the network in a controlled fashion.

We thank Mac Beasley, Terry Orlando, Enrique Tr\'{\i}as, and Shinya
Watanabe for helpful comments.
M.B.\ is grateful for Mac Beasley's hospitality at Stanford.
Research supported through National Science Foundation grant DMS-9500948
(S.H.S.), and by the Ministerio de Educaci\'on y Cultura of Spain (M.B.).


\end{document}